# Underpotential electroless deposition of metals on polyaniline

Amrita Singh[a], Asfiya Contractor[a,*], Ravindra D. Kale[a], Vinay A Juvekar[b]

[a]Department of Fibres and Textile Processing Technology, Institute of Chemical Technology, Mumbai-400019, India.

[b]Department of Chemical Engineering, IIT Bombay, Mumbai 400076, India.

*Corresponding Author:

Email address: asfiyacontractor@gmail.com

**Abstract:** A novel technique to deposit metals on highly conjugated polyaniline films has been developed. In general, electrodeposition of metals, having low reduction potential, from aqueous solution, is difficult due to disruptive effect of hydrogen which evolves during the process. This difficulty is avoided using conducting polymers films with high surface mass density. The polymer chains of these films possess a high degree of conjugation. Such a polymer produces highly stable polarons and therefore has the ability to perform underpotential deposition. Our method involves reduction of polyaniline film with formic acid followed by dipping the coated electrode in the metal salt solution. Deposition of the metal is monitored by rise in the open circuit potential of the electrode. Deposition of metals with high surface mass density has been achieved. The metal is most likely present in the polymer as a coordination complex with amine nitrogen. Such form of metal is expected to have higher catalytic activity than the zero-valent metal. We have been able to deposit metals such as Mn and Cu. Among these, Mn cannot be deposited on polymer by any other method.

# 1. Introduction

Metal deposition on a non-conducting substrate is performed for several reasons. Metal can improve conductivity of the substrate for electronic applications such as printed circuit boards, smart textiles, sensors etc. [1,2]. Metal can improve capability of the substrate to absorb electromagnetic waves in flexible EMI (electromagnetic interference) shields[3], metal can also improve electrocatalytic activity of the substrate[4]. The other applications are catalysis, decorative coating, conducting inks etc.

The most frequently used technique to deposit metals on non-conducting substrates is electroless deposition, which involves sensitization (using stannous salt) and activation (using palladium) prior to metal deposition. Moreover, in order to avoid precipitation of the metal in the bath, metal complexing agents are employed[5,6]. Therefore, this process suffers from disadvantages such as high cost of reagents and poor bath stability. One way to eliminate these drawbacks is to use conducting polymer as a redox mediator. In this technique, the conducting polymer is coated on the substrate and the metal is deposited on the coated substrate. There are two modes by which polymer can facilitate the deposition process. Polymer can provide conducting pathways for transport of electrons from the oxidation sites to the reduction sites. This accelerates the deposition process. However, it neither eliminates the use of the catalyst on which the oxidation of the reducing agent occurs nor the bath in which both the reductant and the metal ions are present[7, 8, 9]. There is another mode by which the conducting polymer can act as a redox mediator. In this mode, we can view the polymer as a capacitor, which undergoes charging and discharging processes. Here, the solutions of the reductant and the metal salt can be segregated. When the polymer is immersed in the solution of the reductant, it undergoes reduction, which is equivalent to the electrical discharge of a capacitor. During this process, the electric potential of the polymer decreases with time. When the polymer in its electrically discharged state is dipped into the solution of a suitable metal salt, it can perform autogenous reduction of the metal ions and itself gets electrically charged during the process. There are several advantages of this technique over the conventional one. First, it eliminates the use of the sensitizer and the activator. Second, it eliminates the use of the conventional electroless bath and, therefore, the associated instability problems.

In the present work, we have explored the second mode of deposition of metals using polyaniline as a capacitor. We begin with the polyaniline film deposited on platinum electrode. The film is reduced to the protonated leucoemeraldine state using formic acid as the reducing agent. In the second step, the electrode in the reduced state is dipped into the solution of the metal to be deposited in acid medium. Here, the metal ions are reduced and the polyaniline is oxidized to the protonated emeraldine state. We have monitored the open circuit potential of the electrode during both reduction and the oxidation of polyaniline electrode. The quantity of the metal deposited on the electrode was also measured.

Our most important finding is that, using this technique, it is possible to deposit metals such as manganese, iron and cobalt,which have standard reduction potentials well below the potential at which the deposition occurs. Hence, we have termed this deposition as underpotential deposition. Of course, the deposited metal is not expected to be in its zero-valent state, but in some intermediate valence state, which is stabilized through its interaction with polyaniline. We have found that the key factor which promotes the ability of polyaniline to perform the underpotential metal deposition is directly related to stability of polarons (radical cations) present in the polymer film. Stability of polarons increases with increase in the degree of conjugation (due to delocalization of the positive charge associated with polaronic nitrogen). The degree of conjugation itself increases with increase in the length of the polymer chain, which is directly related to the surface mass density of the polymer on the electrode. We have shown that no significant deposition of any metals occur unless the surface mass density of the polymer exceeds a critical magnitude (with the exception of ferrous ions). Beyond this mass density, there is a rapid increase in the mass of the metal deposit with the surface mass density of the polymer.

2. Materials and methods

Aniline, high purity sulfuric acid (98%w/w), nitric acid (65% w/w), butyl acetate, hydrochloric acid (37%) and the metal salts,viz. copper(II)sulphate pentahydrate, manganese(II)sulphate monohydrate and ammonium iron (II) sulphate hexahydrate were purchased from Merck (India). Sodium diethyldithiocarbamate and polyaniline powder (15000 Da) were purchased from Sigma-Aldrich(Mumbai, India). Potassium persulfate for analytical purpose was obtained from SD Fine Chemicals(Mumbai, India).All chemicals were used as such except aniline which was distilled to a colourless liquid before use. Also, all dilutions were performed using Milli-Q water.

The setup comprised of a three-electrode cell, where saturated calomel electrode was used as thereference electrode. All potentials reported here are measured with respect to SCE.

The working electrode was a rectangular platinum plate having dimensions 40.7 mm x 10.5 mm x 0.7 mm (geometric surface area 920 mm$^2$). The counter electrode was a cylindrical platinum gauze of large surface area. Electrochemical workstation (CH Instruments Model 600D) was used to control the electrical inputs and outputs. UV-VIS 160(Shimadzu,Japan) was used for colorimetric analysis.

Platinum plate electrode was used for electro-deposition of polyaniline film.The plate was pre-treated with chromic acid, polished with alumina paste and activated in 0.5 M sulphuric acid before use. The electrodeposition of polyaniline was performed using cyclic voltammetry at the scan rate of 50 mV.s$^{-1}$, in a 100 mL aqueous solution containing 0.5 M sulphuric acid and 0.1 M aniline. Deposition was performed in two steps.The first step was conducted in the range from -0.2 to 1V under stirring for 10 cycles and the second step from -0.2 to 0.75V without stirring[10]. At the end of the deposition,the plate was washed with water and further subjected to cyclic voltammetry in 0.5 M sulphuric acid under stirring for ten cycles, at the scan rate of 50 mV.s$^{-1}$. The purpose of the last step was to disengage loose polyaniline chains from the film. The voltammogram of the last cycle of this step was recorded and was used as the characteristic of the film for further analysis. The mass density of polyaniline (mass per unit area of the plate) was varied from 2g.m$^{-2}$ to 70 g.m$^{-2}$. This corresponds to variation of the film thickness from 1.47μm to 51.47μm (considering the density of polyaniline of 1360kg.m$^{-3}$).

Mass of the polymer deposited on the electrode was determined by dissolving it in 20 mL of concentrated nitric acid. The solution was then diluted to 500 mL and the absorbance of the solution was measured at wavelength of 345nm(which corresponds to the benzenoid $\pi \rightarrow \pi^*$ transition). A calibration chart was prepared by measuring the absorbance of different concentration of polyaniline powder (purchased from Sigma-Aldrich) in concentrated nitric acid. Validity of this calibration was checked as follows. The electrodeposited film was scraped from the plate and weighed. It was then dissolved, diluted, and its UV response was measured. The mass of the polymer, estimated from the previously prepared calibration chart tallied with the actual mass of the film within 3% error.

Before the metal deposition, the film is reduced using 0.3 M formic acid solution under stirring. The open circuit potential of the electrode was monitored during the reduction. The process was terminated when the potential of the film reached nearly constant value. In the case of thin films, the potential at the end of reduction step was about -0.2V while as films thickened, the potential after reduction levelled off at higher values (-0.06V).After reduction of the film, the plate was rinsed with water and then subjected to metal deposition.

For the purpose of metal deposition, the plate carrying the reduced polyaniline film was immersed in 100 mL solution containing 0.1 M metal saltin 0.5M sulphuric acid, kept under stirring for about 16 hours. The open circuit potential was recorded over a short period at the beginning and at the end of the process. After deposition, the film was rinsed with water. The quantity of the metal deposited in the polymer was estimated by electrochemically dissolving the metal in a known quantity of aqueous acid and estimating the concentration of the dissolved metal in the solution. Metal dissolution was conducted using cyclic voltammetry in 100 mL of 0.5 M sulphuric acid under stirring. The range of the potential was -0.2 to 0.75V and the scan rate was 10 mV.s$^{-1}$. A total of eleven cycles were performed. In some experiments, chronoamperometry and chronopotentiometry was used for stripping to find the fraction of the metal retained in the film as a function of the electrode potential.

For quantification of managanese, the solution was oxidised to potassium permanganate using potassium persulfate and analysed spectrophotometrically at wavelength of 525nm. For quantification of $Cu^{+2}$ions, the solution was first neutralized using ammonia followed by addition of diethyldithiocarbamate to form the corresponding metal complex.The complex was extracted in butyl acetate and spectrophotometrically analysed at wavelength of 435nm(Cu).The details of the analytical techniques can befound in[11]. In each case, the blank used in the spectrophotometer was prepared by subjecting the stripping solution to the same reagent treatment as was applied to the sample.

### 3. Results and discussion

Since the ability of polyaniline to deposit some metals was found to increase significantly beyond certain surface mass density of the polymer, we have synthesized the polyaniline films with high surface mass density. These films possess high degrees of conjugation and their current-potential characteristics differ significantly compared to thin films. Since these characteristics are helpful in understanding themetal-deposition behaviour, we havedescribed them in details here.

Figure-1 (a) to (c) show cyclic voltammogramsand cyclic coulogramsof three polyaniline films withincreasing mass density.It is clear from these figures that as we increase the mass density of polyaniline, the features of the voltammograms smoothen. The three familiar peaks of polyaniline, which are clearly visible in the lowest mass density film, are subdued at the intermediate mass density and completely vanish at the highest mass density. In the last case, the voltammogram attains a leaf-like shape (Figure c). This shape is indicative of merging of the energy bands and reduction in the spacing between consecutive energy levels allowing almost continuous transition from one energy level to the next.

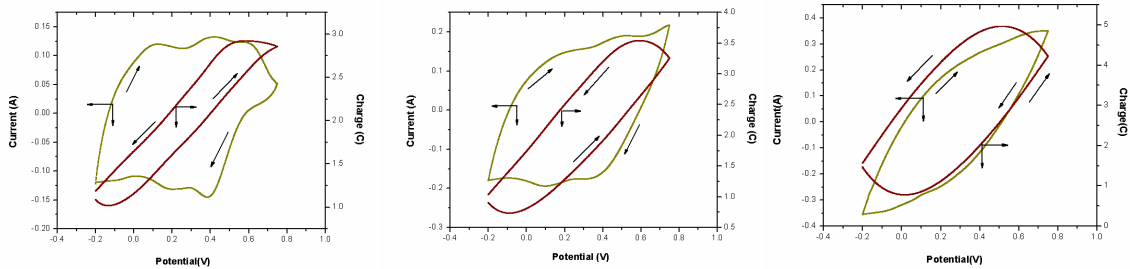

**Figure 1: Cyclic voltammogram with inset coulogram plot of film of different surface mass densities. Concentration of $H_2SO_4$= 0.5 M, concentration of aniline= 0.1 M and temperature = $25^0$ C. Curve a: 5.55 g.m$^{-2}$, curve b: 8.706 g.m$^{-2}$ and curve c: 26.4 g.m$^{-2}$. All potentials are reported with repect to SCE.**

The specific differential capacitance of the film, $c_f$, can be estimated from the coulograms using the equation,

$$c_f = (d\sigma/dE) \tag{1}$$

where, $\sigma$ is the surface charge density, and $E$, the electrode potential. It is seen from the coulograms that within the potential range relevant to metal deposition (-0.2 to 0.4 $V_{SCE}$), the charge density is a linear function of the potential. This indicates that the differential capacitance of the film is constant in this range. Constant capacitance is a sign of uniform spacing of energy levels. Since, the capacitance is a function of the scan rate, we call this a dynamic capacitance, to distinguish it from the equilibrium capacitance.

In Figure-2, we have plotted $c_f$, at the scan rate $v = 50$ mV.s$^{-1}$ and in the potential range -0.2 to 0.4 $V_{SCE}$, against the surface mass density of the polyaniline film. It is seen from this figure that the capacitance varies linearly with mass density at lower density, but attains a constant value at high mass density. At low mass density, the number of chains per unit area of electrode increases with increase in the mass density. But beyond a certain mass density, number of chains becomes constant due to saturation of the adsorption sites on the electrode surface and any further increase in the mass density results in lengthening of chains at constant number density. Beyond this limit, linear charge density per chain increases with increase in the electrode potential. We expect this to cause an increase in the repulsive double layer interaction among the chains, with consequent increase in the strain energy of the film.

The strain energy sets the upper limit on the linear charge density. At this limiting point, a constant capacitance is attained.

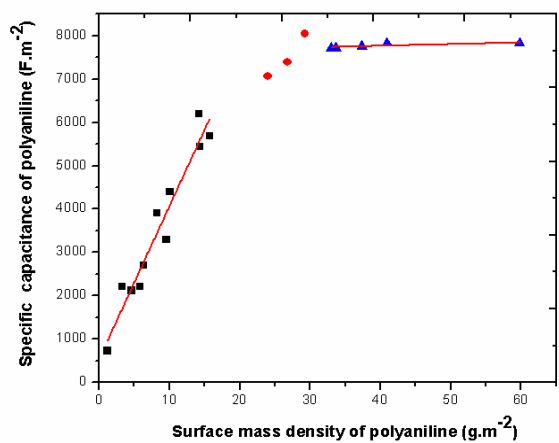

**Figure 2: Specific capacitance versus surface mass density of polyaniline**
**Concentration of sulfuric acid = 0.5 M, scan rate = 50 mV.s$^{-1}$ and temperature = 25$^0$ C.**
**Black squares correspond to liear region. Red circles correspond to intermediate region.**
**Blue triangles correpond to constant region. All potentials are reported with respect to SCE.**

In Figure 3, we have plotted open circuit potential (OCP) of the films as a function of time during its reduction in formic acid. The initial potential of each of the films was different. However, for the sake of clarity, we have included only the portions of OCP below 0 V$_{SCE}$.

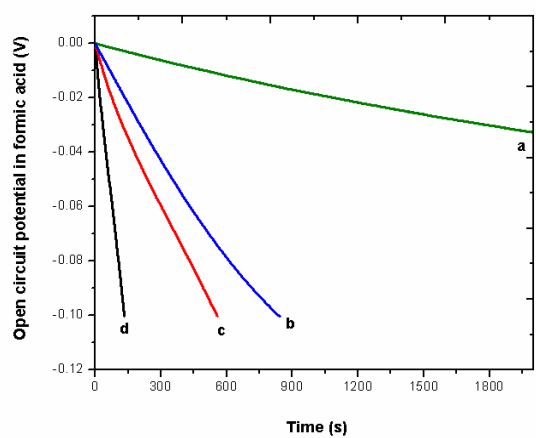

**Figure 3: Open circuit potential ( wrt SCE) versus time plot of polyaniline film.**
**Concentration of HCOOH = 0.3 M and at 25$^0$ C. Each curve corresponds to specific surface mass density.**
**Curve a: 40.6 g.m$^{-2}$, Curve b: 14.33 g.m$^{-2}$, Curve c: 8.23 g.m$^{-2}$ and Curve d: 1.98 g.m$^{-2}$.**

It is seen that OCP reduces with time. This happens due to reduction of polarons( i.e radical cation species $-N^{·+}H-$ ) in the film by formic acid according to the following reactions

$$HCOO^- \rightarrow H^\cdot + CO_2 \quad (2)$$

$$-N^{\cdot+}H- + H^\cdot \rightarrow -NH- + H^+ \quad (3)$$

The reduction current density $i_R$ can be related to rate of change of the open circuit potential ($E_{OCP}$) through the following equation

$$i_R = -\frac{dq}{dt} = -c_f \frac{dE_{OCP}}{dt} \quad (4)$$

It is seen from Figure 3 that there is a linear variation of potential of the film with time in most cases. This indicates that $dE_{OCP}/dt$ can be considered as constant in equation 4 and since for a given film, $c_f$ is also constant, equation 4 indicates that the rate of reduction of the film by formic acid is constant. However, the concentration of polarons is expected to reduce significantly with time in this range of potential. Hence, a constant reduction current implies that the rate of reduction is independent of polaron concentration. We therefore expect that the rate controllong step is breaking of the –C-H bond of formate ( reaction of Eq 2) to generate $H^\cdot$ radicals. These radicals are very active and their reaction with the polarons is very fast. Hence, although the polarons do undergo reduction, the overall reaction rate does not depend on the concentration of polarons.

The rate of reduction has a strong dependence on the surface mass density of the polymer as shown in Figure 4, where we have plotted the reduction current density against the surface mass density. The reduction current density is estimated from the linear portions of the curves in Figure 3, using equation 4.

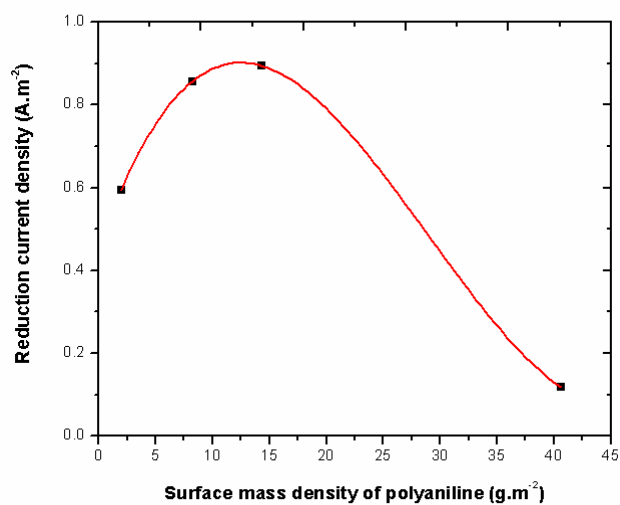

**Figure 4: Reduction current density versus surface mass density of polyaniline film. using 0.3 M HCOOH at $25^0$ C. All potentials are reported with respect to SCE.**

We see that at first, the reduction current density increases with increasing mass density, reaches a peak and then decreases at higher mass densities. This indicates that reaction represented by Eq 2 occurs only when formate ions are in the form of counterions of amine groups i.e. as $(-NH_2^+ -)(OOCH)^-$. With increase in the mass density, the concentration of the amine formate groups, should increase, thereby increasing the rate of reduction. However, the basicity of the amine also decreases with increase in the mass density. This decrease is caused by delocalization of hydrogen radicals on the protonated amine nitrogens on the polymer backbone.

$$(-NH_2^+ -) \rightarrow (-N^{\cdot +}H -) + H^{\cdot} \qquad (5)$$

Decrease in the basicity suppresses the tendency to form $(-NH_2^+ -)(OOCH)^-$, thereby reducing the rate of reduction at very high mass density. It is important to realise that diffusion of ions through the film is not a rate controlling step, since the diffusion time constant is much shorter than the time interval over which the film is reduced. For example, for the film with the surface mass density of 40.6 g.m$^{-2}$, the film thickness is estimated to be about 50 μm. Using the diffustion coefficient of formate ions of the order of 10$^{-10}$ m$^2$.s$^{-1}$, the diffusion time constant is estimated is of the order of 25 s, which is very short compared to the reduction time is of the order of 1000 s.

Deviation from the linear OCP trend is observed at lower potentials as shown in Figure 5, for the thinnest and thickest films at potentials lower than those depicted in Figure 3. We see that the OCP plots not only become curved, but also show reversal in the trend after passing through a minimum. We assign cause of this reversal to oxygen reduction reaction(ORR) which also occurs on the film. It is increasingly dominant at low potentials[12]. Since ORR oxidizes the polymer, it causes the potential to rise. In the case of very thick films, the ORR reaction dominates over formic acid reduction reaction even at higher potentials.

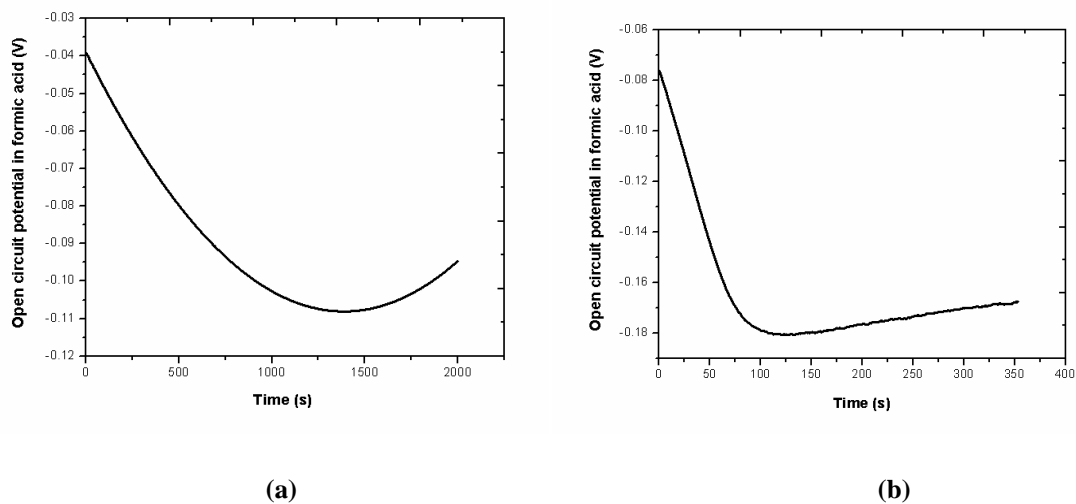

(a)  (b)

**Figure 5: Open circuit potential versus time plot of polyaniline film.
HCOOH = 0.3 M at $25^0$ C. Curve a: 45.5 g.m$^{-2}$ and b: 5.85 g.m$^{-2}$.
All potentials are reported with repect to SCE.**

Since this reversal would cause interference in the present work, we terminated the reduction process as soon as OCP began to reverse the direction.

We now consider the results of deposition of manganese and copper on the reduced polyaniline films. Manganese is one of the most difficult metal to deposit since it has standard reduction potential of $-1.185\ V_{RHE}$. It cannot be deposited by any known electroless deposition technique. We used 0.1 M aqueous solution of manganese sulphate in 0.5 M sulfuric acid for the purpose of deposition. The polyaniline film was reduced in 0.3 M formic acid to the lowest attainable potential and then immersed in the manganese sulphate solution. The progress of the deposition reaction was monitored using the open circuit potential. Figure 6 shows the open circuit potential for two films, one with low mass density and the other with high mass density. For the sake of comparison, we have also plotted the OCP of the same film in 0.5 M sulfuric acid, but in the absence of the manganese salt.

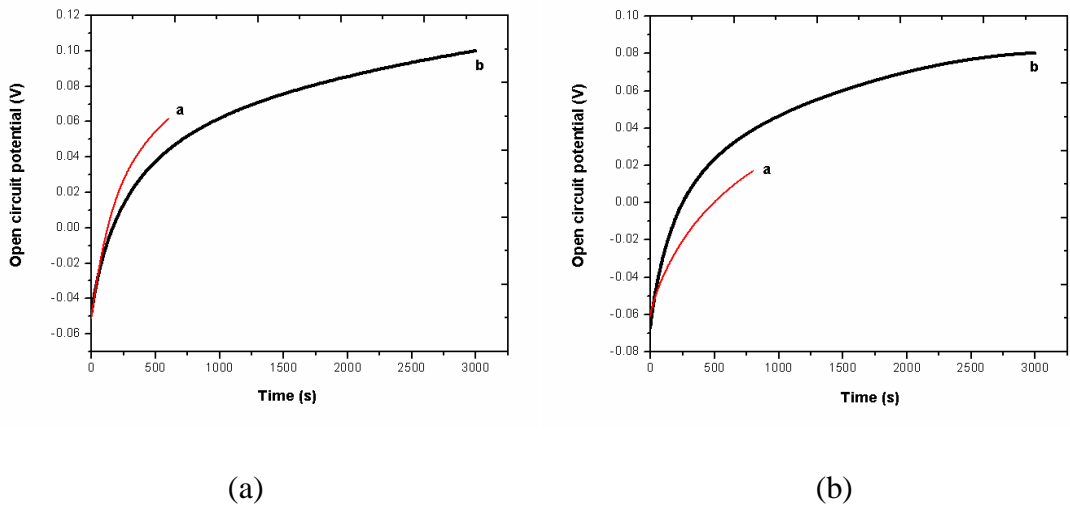

(a)                                                      (b)

**Figure 6: Evolution of open circuit potentials ( wrt SCE) of two polyaniline films ( mass density in Figure 6 (a) is 14.16 g.m$^{-2}$ and for Figure 6 (b) is 28.5 g.m$^{-2}$**
**Curves a in each figure corresponds to OCP in 0.5 M sulfuric acid, and curve b corresponds to OCP in 0.5 M H$_2$SO$_4$ and 0.1 M MnSO$_4$. Temperature: 25$^0$ C.**

We see from the figure that for the low mass density film, the evolution of OCP is slowed down in the presence of managanese salt, whereas for the high mass density film, it is hastened in the presence of the salt. It is important to note that in the presence of sulfuric acid, the potential of the film rises due to ORR, which generates polarons. In the presence of manganese sulphate, both ORR and metal reduction reactions occur simultaneously and both produce polarons which causes rise in the potential of the electrode. These reactions may compete with each other and thereby retard the rate of each other. The overall rate (which is manifested by the rise in OCP) will depend on extent of the mutual retardation. Comparison of the two cases showes that the retardation effect is much stronger in thin films compared to thick films. It is likely that for thick films, the two reactions occured independently since the OCP evolution is faster in the presence of the salt.

We propose the following mechanism to explain the trends in Figure 6. As a first step, managanese ion forms a coordination complex with the amine nitrogen. This complex undergoes internal electron transfer reaction, where one of the electrons on nitrogen gets accommodated in the 4s orbital of manganese and simultaneously the nitrogen atom is converted into a radical cation (polaron). This hinders the protonation of amine, which is a prerequisite for ORR reaction[12]. Hence the OCP evolution slows down in the presence of manganese salt in the case of thin polyaniline films. For thicker films, abundance of amine groups allow the two reactions to occur indepednently.

To gain more understanding of the metal deposition reaction, we conducted an experiment where the polyaniline coated electrode was dipped in the metal solution bath for different time periods and the amount of manganese deposited on the film was estimated. At the end of each period, the electrode was removed from the solution, washed thoroughly and Mn was stripped from the film using chronopotentiometry. The metal-free film was again dipped in same bath for a longer period and the same procedure repeated. We also noted the open circuit potential attained at the end of each period. The film attained the saturation with respect to metal deposit after 16 hours. In Figure 7, we have plotted the fractional deposition of Mn (the ratio of the actual mass to the saturation mass) at different times and at different OCP values.

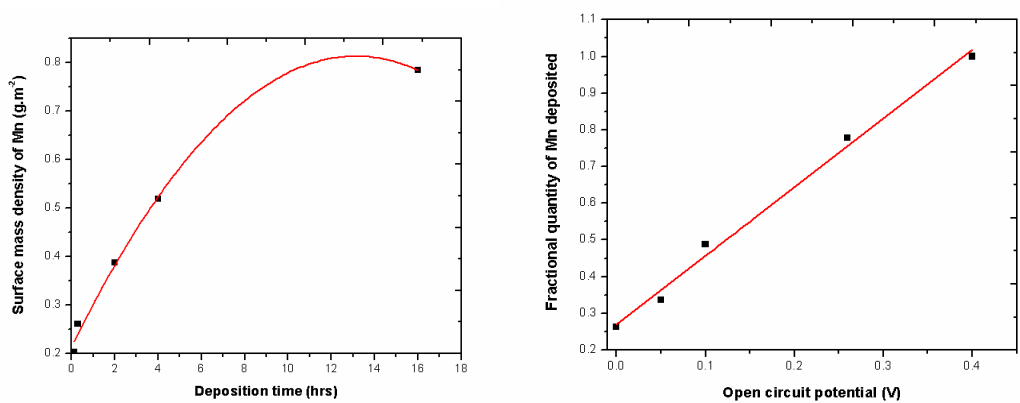

**Figure 7: Surface mass density of manganese versus (a) Deposition time and (b)Open circuit potential uisng 0.1 M manganese salt, 0.5 M sulfuric acid with surface mass density of 30 g.m$^{-2}$ polyaniline film at 25$^0$ C. All potentials are reported with respect to SCE.**

It is seen from this figure that rate of deposition is rapid during the short initial period showing initial jump. This followed a linear regime, where the rate of deposition remains constant. In the last phase, the rate of deposition decreases with time. However, the most important observation is made from Figure 7 (b), where we see that the extent of deposition varies linearly with open circuit potential. Since the capacitance of the film is constant, the concentration of polarons in the film is directly proportional to the open circuit potential of the film. This implies that the mass of manganese deposited in the film is directly proportional to the concentration of polarons. This means that manganese ion forms a complex with the radical cation (polaron) and not with the pi-electrons of aromatic ring. If the latter were true, we would have obtained a weak dependence of metal deposit on potential.

From the capacitance of the film (8000 $F.m^{-2}$) of Figure 8, we can estimate the capacity of generation of polarons per unit area as 0.0829 $mol.m^{-2}.V^{-1}$. The total mass of Mn deposited in the film is 0.314mg at potential of 0.39 V and 0.0785 mg at 0V. This gives the metal deposition capacity as 0.0274 $mol.m^{-2}.V^{-1}$, which is 33% of the maximum capacity of polarons generated in the film per unit area. This means that ORR reaction has a major contribution towards generation of polarons.

To probe this reaction further, we plotted surface mass density of metal (based on the geometric surface area of the electrode) as a function of the surface mass density of the polyaniline(Figure 8a) and also against the differential capacitance of the film (Figure 8b).

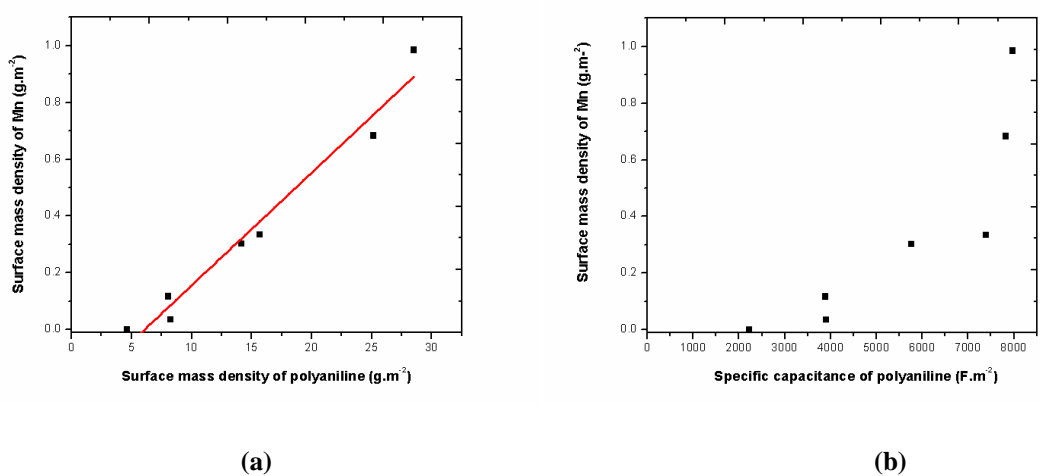

(a)                                                                 (b)

**Figure 8: Surface mass density of manganese using 0.1 M manganese salt and 0.5 M sulfuric acid at $25^0$ C versus (a) surface mass density of polyaniline (b) specific capacitance.
All potentials are reported with respect to SCE**

An important point to be noted from Figure 8 is that, there is a critical mass density, below which no deposition of manganese occurs. This value was found from the above figure as 6 $g.m^{-2}$. In terms of dynamic capacitance it is about 2200 $F.m^{-2}$. Since mass density (or capacitance) of the film is related to the degree of conjugation in the polymer chains, this observation implies that a minimum degree of conjugation is needed for the deposition to commence.

Next, we consider copper deposition. It was conducted on the film which was prereduced in formic acid. Deposition was performed in 0.1 M $CuSO_4$ in 0.5 M sulfuric acid at $25^0$ C. In Figure 9, we have plotted surface mass density of the metal as a function of surface mass density of the polyaniline and specific capacitance of the film.

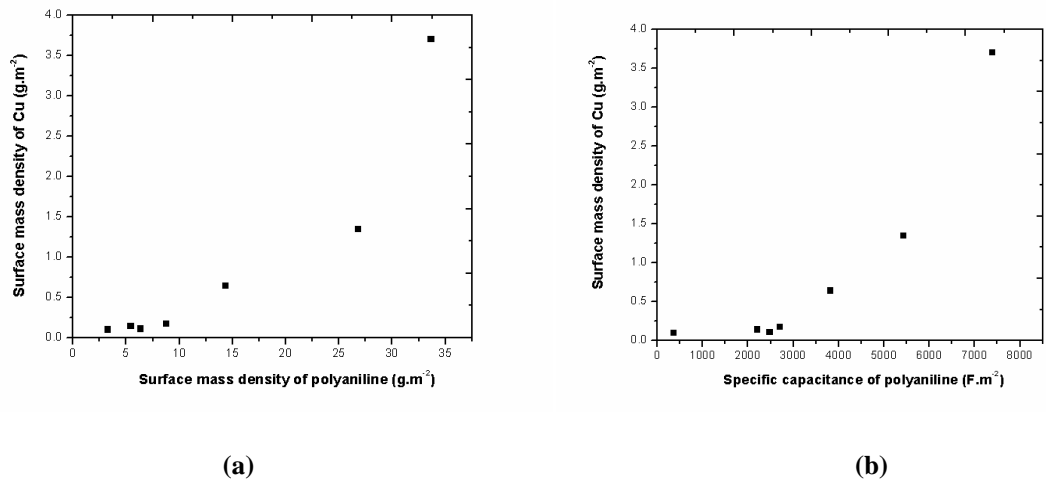

(a)　　　　　　　　　　　　　　　(b)

**Figure 9:** Surface mass density of copper versus (a) surface mass density of polyaniline (b) specific capacitance. All potentials are reported with respect to SCE.

It is seen from the figure that only a small amount copper is deposited unless the critical specific mass (about 8 g.m$^{-2}$) orcapacitance (2000 F.m$^{-2}$) is exceeded. It should be noted that the standard reduction potential for cupric ions to copper is 0.34 V$_{SHE}$ or 0.102 V$_{SCE}$. Since formic acid reduces the potential below 0 V$_{SCE}$, there is a possibility of direct deposition of copper. But the quantity depsited is expected to be small. We can therefore conjecture that up to the polymer mass density below about 8 g.m$^{-2}$, the amount of copper deposited is due to direct reduction of copper ions.

From comparison of Figures 8 and 9, we see that the critical value of both the mass density and capacitance in the case of copper are lower than that for manganese. This is expected since manganese ions are difficult to reduce than copper ions. However, based on similarity in trends we expect copper deposition to follow the same mechanism as deposition of manganese. The only difference is that now the interaction of the amine nitrogen occurs with 3d$^9$ orbital of cupric ion. Initially, nitrogen forms a coordinate complex with cupric ion in which one of the electron of nitrogen is coordinated with the d orbital of copper. This is followed by transfer of electron from nitrogen to form the cuprous ion with ten electrons in 3d orbital. The nitrogen itself gets transformed to a radical cation and forms a valence complex with polaron.

It is instructive to find out what fraction of polarons is occupied by copper present in the film. As an example, we consider the point with highest capacitance (7500F.m$^{-2}$) in Figure 9(b). The potential at the beginning of deposition was -0.06V and that at the end, it was 0.4 V. This gives the total surface charge density associated with polarons of 3450 $C.m^{-2}$. This corresponds to surface density of polarons of 0.0357 $mol.m^{-2}$. The amount of copper deposited was 3.7 $g.m^{-2}$, which corresponds to 0.0587 $mol.m^{-2}$. This quantity is 64% higher than the moles of polarons present. This calculation implies that the film contained more polarons than the estimation based on capacitance of the film measured in sulfuric acid alone. More polarons must be getting generated in the film in the presence of metal. This trend should also be followed by manganese. We could not detect this phenomena in the case of manganese since, the amount of manganese deposited was much smaller than the polaronic charge.

To find how strongly copper binds to the polymer, we performed an experiment in which we stripped copper from the film in 0.5 M sulfuric acid under stirring, using chronoamperometry. We stepped down the potential in four equal steps from potential of 0.46 $V_{SCE}$ down to -0.04 $V_{SCE}$. In Figure 10, we have plotted fraction of the copper retained as a function of potential.

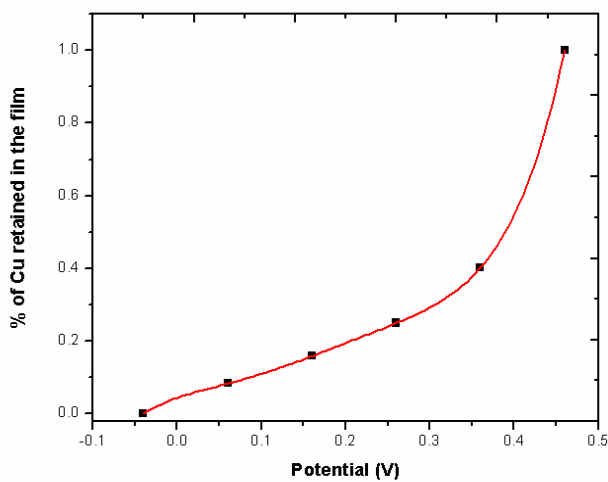

**Figure 10: Percentage of copper retained as a function of potential using 0.1 M copper salt and 0.5 M sulfuric acid and polyaniline 5.5 $g.m^{-2}$ at $25^0$ C. All potentials are reported with respect to SCE**

From figure 10 it is noted that only 40% of the deposited copper is retained in the film at the end of the first step where we stepped down the potential from 0.46 $V_{SCE}$ to 0.36 $V_{SCE}$. This means that 60% of the copper gets stripped in a narrow range at higher potential. The standard reduction potential of copper 0.102$V_{SCE}$. Hence if copper were deposited as metal, all of it would have dissolved at potential of 0.36 $V_{SCE}$. However retention of significant fraction of the copper in the film indicates that it is strongly bound to the polaronic nitrogen.

We therefore propose the following hypothesis for the underpotential deposition of divalent metal ions such as manganese and copper on polyaniline. Metal ion is initially reduced from its divalent state to monovalent state by oxidizing the amine nitrogen to polaron. This process requires that the resulting polarons have very low electrochemical potential. Hence, valence complex can be autogenously generated only at low potentials. Even here, a minimum degree of conjugation is needed in order that the resulting polarons have sufficiently low electrochemical potential to bring about this reaction in an autogenous manner.

## 4. Conclusion

Polyaniline should have a minimal degree of conjugation before it is capable of reducing metal from its salt. For electrochemically synthesized polymer, conjugation increases with increase in the surface mass density which in turn increases the length of the polymer chains, which is reflected in the film thickness and capacitance. In this work, polymer films having thickness varying from 1.47μm to 51.47 μm and capacitance varying from 800 to 8000 $F.m^{-2}$ have been used for metal deposition.

With increase in the surface mass density beyond the critical value, two regimes are observed. In the first regime, the surface mass density of the deposited metal increases linearly while in the second regime, it increases steeply with increase in the surface mass density of polyaniline film. Moreover, the metal, deposited on the film, exists in the form of a valence complex with the polarons.

Using the underpotential deposition technique described in this work, it is possible to deposit a metal such as Mn having very low standard reduction potential. Since in the coordination complex form, transition metal is likely to possess catalytic activity which is quite different from the zero valent metal, this method can be explored for preparation of catalysts.

## 5. Acknowledgement

The authors wish to acknowledge Science and Engineering Research Board (SERB), Government of India for financial support in carrying out this work.